\documentclass{nature}

\usepackage{amssymb}
\usepackage{graphicx}

\title{The awakening of a classical nova from hibernation}

\author{Przemek Mr\'oz$^{1}$\footnote{Corresponding author.}, Andrzej Udalski$^1$, Pawe\l{} Pietrukowicz$^1$, Micha\l{}~K. Szyma\'nski$^1$, Igor Soszy\'nski$^1$, \L{}ukasz Wyrzykowski$^1$, Rados\l{}aw~Poleski$^{1,2}$, Szymon Koz\l{}owski$^1$, Jan Skowron$^1$, Krzysztof Ulaczyk$^{1,3}$, Dorota Skowron$^1$ \& Micha\l{} Pawlak$^1$}

\begin{document}

\maketitle

\begin{affiliations}
 \item Warsaw University Observatory, Al. Ujazdowskie 4, 00-478 Warsaw, Poland
 \item Department of Astronomy, Ohio State University, 140 W. 18th Ave., Columbus, OH 43210, USA
 \item Department of Physics, University of Warwick, Coventry CV4 7AL, UK
\end{affiliations}

\begin{abstract}
Cataclysmic variable stars -- novae, dwarf novae, and nova-likes -- are close binary systems consisting of a white dwarf star (the primary) that is accreting matter from a low-­mass companion star (the secondary)\cite{warner}. 
From time to time such systems undergo large-amplitude brightenings. The most spectacular eruptions, with a ten-thousandfold increase in brightness, occur in classical novae and are caused by a thermonuclear runaway on the surface of the white dwarf\cite{pacz}. Such eruptions are thought to recur on timescales of ten thousand to a million years\cite{ford}. In­ between, the system's properties depend primarily on the mass­-transfer rate: if it is lower than a billionth of a solar mass per year, the accretion becomes unstable and the matter is dumped onto the white dwarf during quasi­-periodic dwarf nova outbursts\cite{osaki}.
The hibernation hypothesis\cite{shara} predicts that nova eruptions strongly affect the mass-transfer rate in the binary, keeping it high for centuries after the event\cite{kovetz}. Subsequently, the mass-transfer rate should significantly decrease for a thousand to a million years, starting the hibernation phase. After that the nova awakes again -- with accretion returning to the pre-­eruption level and leading to a new nova explosion. The hibernation model predicts cyclical evolution of cataclysmic variables through phases of high and low mass-­transfer. The theory gained some support from the discovery of ancient nova shells around dwarf novae Z Camelopardalis\cite{shara_07} and AT Cancri\cite{shara_12}, but direct evidence for considerable mass­-transfer changes prior, during and after nova­ eruptions has not hitherto been found.
Here we report long­-term observations of the classical nova V1213 Cen (Nova Centauri 2009) covering its pre- and post-­eruption phases and precisely documenting its evolution. Within the six years before the explosion, the system revealed dwarf ­nova outbursts indicative of a low mass­-transfer rate. The post­-nova is two orders of magnitude brighter than the pre-­nova at minimum light with no trace of dwarf nova behaviour, implying that the mass-­transfer rate increased considerably as a result of the nova explosion.
\end{abstract}

\begin{figure}
\includegraphics[width=0.9\textwidth]{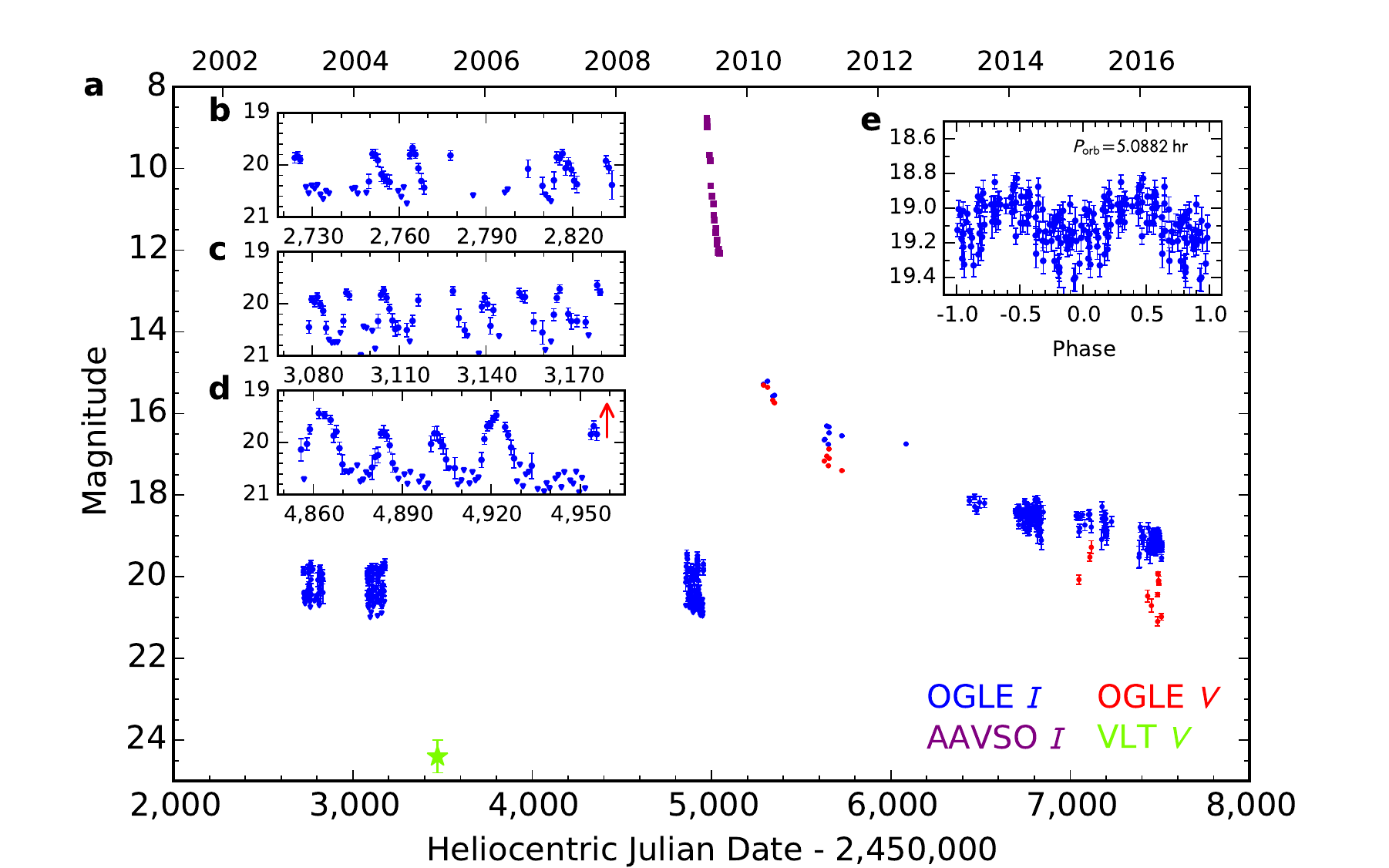}
\caption{\textbf{Long-term light curve of V1213 Cen (Nova Cen 2009). }
\textbf{a,} Blue and red data points are OGLE $I$- and $V$-band observations, respectively. Purple data points were taken from the American Association of Variable Star Observers (AAVSO) database. The green asterisk shows the $V$-band brightness from the deep VLT/VIMOS image.
\textbf{b-d,} Close-up (these panels have the same axes labels as \textbf{a} of the 2003, 2004, and 2009 data, showing dwarf nova outbursts. In quiescence the photometry is limited by the proximity of a bright star, so the dwarf nova must have been fainter than $I=20.8$ mag (Methods). Thus, some observatios are plotted with triangles as upper limits on brightness. The red arrow in \textbf{d} marks the beginning of the nova eruption, which probably started during the last dwarf nova outburst.
\textbf{e,} Post-nova shows variability with a period of 5.1 h, which we interpret as a signature of the orbital period.
Error bars are $1\sigma$.
}
\label{lc}
\end{figure}

\begin{figure}
\centering
\includegraphics[width=0.58\textwidth]{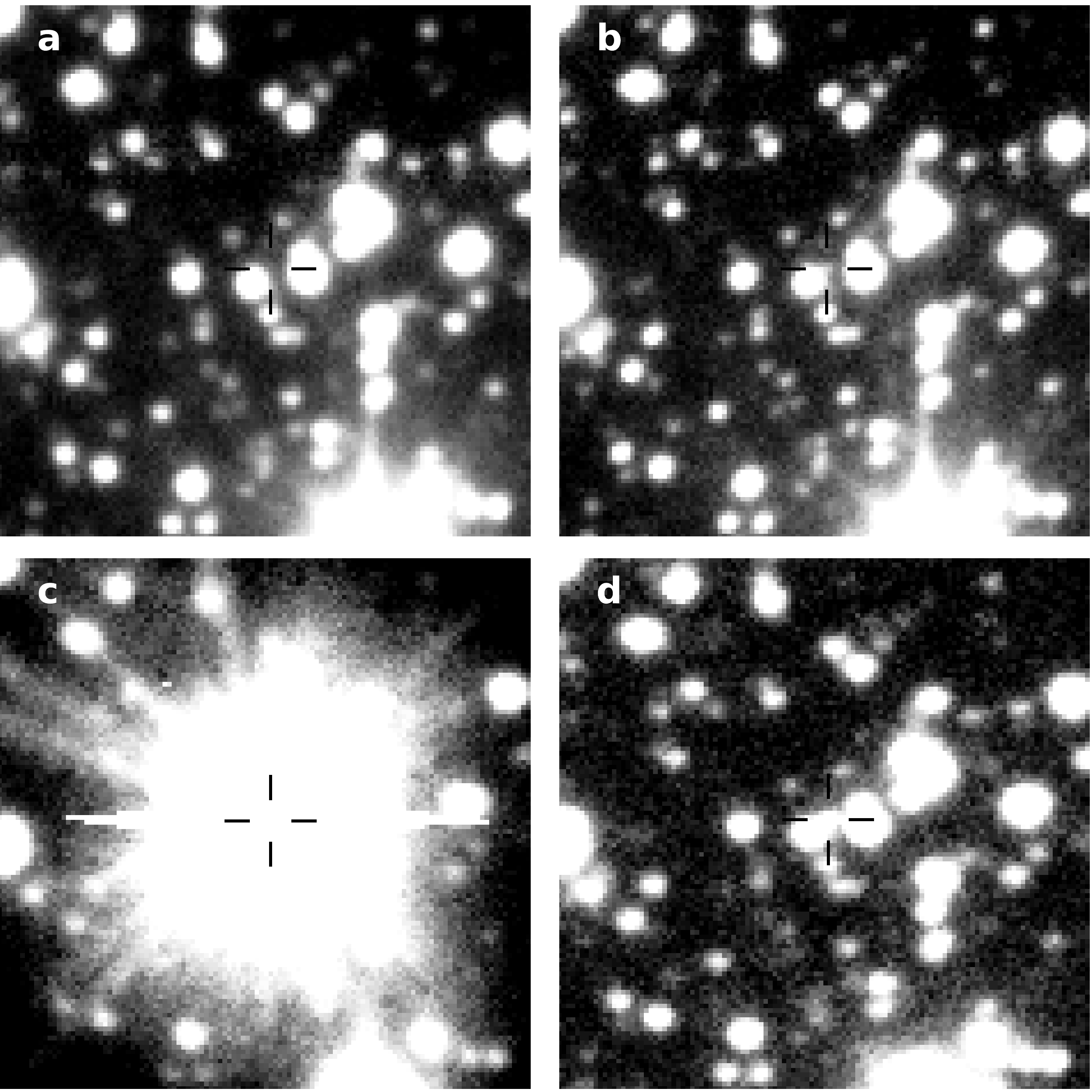}
\caption{\textbf{Snapshots of a nova lifecycle.}
\textbf{a,} Novae spend most of their time as faint, low mass-transfer-rate systems. The pre-nova in quiescence had $V=24.4 \pm 0.4$ mag and it was invisible in this OGLE image from 28 April 2009.
\textbf{b,} The accretion disk's instability leads to dwarf nova outbursts, when the matter is dumped onto the white dwarf. In this image from 4 April 2009, the star was $I\approx 19.9$ mag.
\textbf{c,} The OGLE image from 24 May 2009, acquired 16 days after the nova discovery. The nova is heavily saturated. The nova eruption causes the companion star to lose matter at a much higher rate, probably owing to extreme irradiation.
\textbf{d,} The $I$-band OGLE reference image (2014--2015) showing a bright post-nova. The irradiation-induced high mass-transfer rate is believed to last hundreds of years after the eruption.
All images are $30'' \times 30''$, North is up and East is to the left.
}
\label{fchart}
\end{figure}

The classical nova V1213 Cen (Nova Centauri 2009) was discovered in the image acquired on 2009 May 8.235 UT\cite{pojman1} by the All-Sky Automated Survey\cite{pojman2}. The nova peaked at $V\approx 8.5$ mag and declined by 2 mag in $t_2 = 7.5 \pm 0.5$ days, which corresponds to the absolute brightness of $M_V = -8.5 \pm 0.5$ mag\cite{shafter}$^{\rm ,}$\cite{shafter2}. We estimate the interstellar reddening towards the nova $E(B-V)=0.9 \pm 0.1$ mag and its distance of $7.2^{+2.4}_{-1.8}$ kpc (see Methods), which places the nova in the Scutum-Centaurus Arm of the Milky Way. Both spectroscopic\cite{pigulski}$^{\rm ,}$\cite{walter} and X-ray observations\cite{schwarz_atel}$^{\rm ,}$\cite{schwarz} confirmed that it was a typical example of a fast nova.
The nova was regularly monitored in the course of the Optical Gravitational Lensing Experiment (OGLE) sky survey\cite{udalski} during the years 2003 -- 2016, which allowed us to trace its full photometric history (Figs \ref{lc} and \ref{fchart}). 

Observations from 2010--2016 show a typical slow decline after the nova eruption (Fig. \ref{lc}a). Currently, the star is fading at a rate of $0.48 \pm 0.37$ mag/yr in the $V$-band. The post-nova shows additional sinusoidal variability (Fig. \ref{lc}e) on an orbital period of $5.0882 \pm 0.0017$ h (Methods), which is characteristic for many old novae\cite{mroz}. This suggests that the secondary is a main-sequence star with a mass of about $0.5\ {\rm M_{\odot}}$ (where $\rm M_{\odot}$ is the mass of the Sun).

Pre-eruption data are much more interesting (Fig. \ref{lc}a-d), as these parts of nova light curves have rarely been observed. Reliable photometry from archival photographic plates is available for only a dozen old novae and most of them have the same brightness before and after the eruption\cite{robinson}$^{\rm ,}$\cite{collazi}. Our data for V1213 Cen reveal a characteristic variability, which we interpret as dwarf nova outbursts (see Fig. \ref{lc}b-d). Outbursts peak at $I\approx 19.5$ mag and have an amplitude of $\Delta V \approx 3.0$ mag (assuming an intrinsic colour of $(V-I)_0 = 0.5$ mag\cite{mroz_dn}). They last from five to twelve days (with a mean of six days) with inter-outburst intervals of 12 to 24 days (the mean duty cycle is $38 \pm 5$\% with annual variations from 31 to 44\%). Its outburst properties are typical of a U Geminorum-like dwarf nova\cite{warner} and are consistent with a 5.1-h orbital period (see also Methods and Extended Data Fig. 1).

In quiescence (between outbursts) the system faded to a magnitude of $V \approx 24.4 \pm 0.4$ mag (Methods). This means the star brightened by $\Delta V = 24.4 - 8.5 = 15.9$ mag (over two million times) during the nova eruption and, assuming an absolute brightness during the nova peak of $-8.5 \pm 0.5$ mag, the $V$-band absolute magnitude in the quiescence was $+7.4 \pm 0.7$ mag. Again, this is consistent with the typical luminosities of dwarf novae (see Extended Data Fig. 2). The nova eruption occurred within six days after the start of the last dwarf nova outburst (Fig. \ref{lc}d) and so, although the light curve is incomplete, it is tempting to suppose that matter dumped onto the white dwarf during that last few days of dwarf nova outbursts triggered the thermonuclear runaway and nova explosion.

The quiescent absolute brightness of the dwarf nova corresponds to a mass-transfer rate of about $1 \cdot 10^{-10}\ {\rm M_{\odot}/yr}$\cite{smak}. At such a low mass-transfer rate the accretion disk is in an unstable state\cite{warner}, which is consistent with the presence of dwarf nova outbursts before the 2009 eruption. The accretion disk produces most of the light in the $V$-band, so we can estimate how the mass-transfer rate changed by simply comparing pre- and post-eruption $V$-band brightness. At present in 2016, the system has $V=20.10 \pm 0.03$ mag ($V=17.14 \pm 0.08$ mag in 2011), which means the mass-transfer rate is now $10^{0.4\cdot(24.4-20.1)}=52$ times higher than before the nova eruption (in 2011 it was 800 times brighter). The absolute brightness of the star is $M_V=+3.1 \pm 0.5$ mag (see Extended Data Fig. 2). Although the exact cause of the mass-transfer rate increase remains unknown, its magnitude is consistent with irradiation of the secondary\cite{kovetz} rather than the interaction between nova ejecta and a companion star (unless the eruption was very asymmetric)\cite{schreiber}$^{\rm ,}$\cite{nelemans}.

Most known pre- and post-novae have the same brightness\cite{robinson}$^{\rm ,}$\cite{collazi} and hence the same mass-transfer rate. We asked why V1213 Cen exploded as a classical nova during the dwarf nova stage. V1213~Cen was a fast nova, so the white dwarf should be massive ($\gtrsim 1{\rm M_{\odot}}$) and the ignition mass should have been relatively small ($\approx 10^{-6}-10^{-5} {\rm M_{\odot}}$)\cite{prialnik}. The mass that accumulated immediately after the previous nova eruption (and before the start of hibernation) must have been close to the ignition mass, which is consistent with the theoretical predictions for the duration of the irradiation-induced high mass-transfer rate ($\approx 10^2$ yrs)\cite{kovetz}. This timescale suggests a natural definition of recurrent novae (a small group of novae for which at least two eruptions have been recorded) as novae that have inter-eruption times short enough that they do not fall into hibernation. Another possibility is that the 2009 eruption was the first nova explosion in this system; however, we consider this hypothesis unlikely because most cataclysmic variables start their evolution at longer orbital periods. 

V1213 Cen is now slowly fading. What will be its fate? One can expect that the system will remain bright for a few decades and then it will again transform into a dwarf nova, following the hibernation theory predictions. There are four known cases of post-novae that showed dwarf-nova-like outbursts. V446 Her (Nova Herculis 1960) is the system most resembling V1213 Cen. It has a similar orbital period of 4.97 h\cite{thorstensen}. Since the 1990s it has been showing low-amplitude outbursts\cite{honeycutt} similar to those observed in other long-period (3--10 hr) dwarf novae. On the other hand, two other novae -- GK Per (Nova Persei 1901) and V1017 Sgr (Nova Sagittari 1919) -- exhibit regular dwarf nova outbursts with amplitudes of 1-3 mag and recurrence times of years. Their outbursts are much longer than in other dwarf novae, because their orbital periods are unusually long (1.99 and 5.7 days, respectively), indicating that the secondary is an evolved star\cite{sekiguchi}. Therefore, they are not typical cataclysmic binaries (with low-mass main-sequence secondaries). Finally, BK Lyn, which has been proposed to be the Chinese Nova of 101 AD has, recently undergone the transition from a nova-like to a dwarf nova\cite{patterson}, which can be interpreted as a sign of hibernation\cite{shara14}. BK Lyn is a short-period system (1.80 h), so the irradiation effect might be stronger there than in long-period novae. Thus, the time required to the switch was considerably longer ($\sim2000$ years) for BK Lyn than for the other three cases.

V1213 Cen with its well known pre- and post-eruption behaviour can become a Rosetta stone for nova evolution studies. Its extensive follow-up observations in the next decades will enable further tests of the long-term nova evolution.

\begin{addendum}
 \item[Acknowledgements] We would like to thank Profs. M. Kubiak and G. Pietrzy{\'n}ski, former members of the OGLE team, for their contribution to the collection of the OGLE photometric data over the past years. P.M. is supported by the ``Diamond Grant'' No. DI2013/014743 funded by the Polish Ministry of Science and Higher Education. The OGLE project has received funding from the National Science Center, Poland, grant MAESTRO 2014/14/A/ST9/00121 to A.U. We acknowledge the variable star observations from the AAVSO International Database contributed by observers worldwide and used in this research.
 \item[Author Contributions] P.M. analysed and interpreted the data, and prepared the manuscript. A.U. reduced and analysed the OGLE photometry. P.P. analysed the VLT data. All authors collected the OGLE photometric observations, reviewed, discussed and commented on the present results and on the manuscript.
 \item[Author Information] The time-series photometry of V1213 Cen is available to the astronomical community from the OGLE Internet Archive (ftp://ftp.astrouw.edu.pl/ogle/ogle4/V1213Cen). Reprints and permissions information is available at www.nature.com/reprints. The authors declare that they have no competing financial interests. Readers are welcome to comment on the online version of the paper. Correspondence and requests for materials should be addressed to P.M. (pmroz@astrouw.edu.pl).
\end{addendum}

\newpage

\newpage

\begin{methods}

\subsection{Reddening and distance.}

The intrinsic colour of novae at maximum light\cite{vdb} is $(B-V)_0=+0.23 \pm 0.06$ mag and 2 mag below the peak magnitude it is $(B-V)_0 = -0.02 \pm 0.04$ mag. The first multicolour photometry of V1213 Cen was collected by amateur astronomers from the American Association of Variable Star Observers (AAVSO)\cite{kafka} five days after the peak. However, the data from $t_2 \pm 1$ day show considerable scatter between $(B-V)=0.12$ and $0.61$ mag, much larger than the quoted error bars ($\sim$ one hundredth of a magnitude), plausibly meaning that they suffer from unknown systematic effects. AAVSO measurements are differential, that is, they show relative difference in magnitudes between the nova and one or more (``ensemble'') comparison stars. We consider the latter to be more reliable. The available data were acquired by two observers, L. Elenin (ELE) and N. Butterworth (BIW), and cluster around $(B-V)\sim 0.4$ or $\sim 0.7$ mag, respectively. That shows that at least one dataset is affected by some systematic error(s) and hence the reddening $E(B-V)>0.4$ mag.

The {\it Swift} X-ray spectrum of the nova can be fitted\cite{schwarz_atel} as a blackbody with temperature $kT=31 \pm 4$ eV and an absorbing column density $N_{\rm H} = (6.4 \pm 1.8) \cdot 10^{21}$ cm$^{-2}$, which corresponds\cite{guver} to a $V$-band extinction of $A_V = 2.9 \pm 0.9$ mag (or $E(B-V)=1.0 \pm 0.3$ mag assuming the standard extinction law with $R_V=A_V/E(B-V) = 3.1$). Since the blackbody model is an oversimplification of the real X-ray spectrum and the measured absorbing column density may contain the internal absorption, the interstellar reddening $E(B-V)<1.0$ mag.

Using the multicolour light curves of several novae, we estimated the intrinsic colour of novae two mag below the peak magnitude of $(V-I)_0 = 0.62 \pm 0.17$ mag (three mag below the peak brightness of $(V-I)_0 = 0.53 \pm 0.23$ mag). The AAVSO $(V-I)$ colour curve of V1213 Cen shows much less scatter than in $(B-V)$ and hence we were able to assess $E(V-I)=1.39 \pm 0.15$ mag, which corresponds to $E(B-V)=0.90 \pm 0.10$ mag.

Hence, we assume that the reddening is $E(B-V)=0.9 \pm 0.1$ mag. If the reddening law toward the nova is standard, then $A_V = 2.7 \pm 0.3$ mag and the distance modulus of the nova is $14.3 \pm 0.6$ mag (distance $7.2^{+2.4}_{-1.6}$ kpc). This places the nova in the Scutum-Centaurus Arm of the Milky Way. At present in 2016, the observed colour of the nova is $(V-I)=1.14 \pm 0.03$ mag, so the estimated colour excess $E(V-I)=1.4 \pm 0.2$ mag is consistent with intrinsic colours of post-novae.

\subsection{OGLE data.}

The OGLE data were collected during two phases of the project: OGLE-III\cite{udalski2008} (2001-2009) and OGLE-IV\cite{udalski} (since 2010). The survey uses the dedicated 1.3-m Warsaw Telescope located at Las Campanas Observatory, Chile, operated by the Carnegie Institution for Science. All observations were taken in the $I$ and $V$ filters, closely resembling the standard system. Data were reduced, and astrometrically and photometrically calibrated using the standard pipelines\cite{udalski}$^{\rm ,}$\cite{udalski2008}$^{\rm ,}$\cite{szymanski}. Thanks to the superb quality of images and application of the Difference Image Analysis technique\cite{wozniak}, the photometry is very accurate (3\% for 18 mag, 10\% for 19.5 mag), although the nova has close, bright neighbours (Fig. \ref{fchart}).

Pre-eruption images were collected during the OGLE-III phase between March 2003 and May 2009 (2374 $I$-band images with an exposure time 180~s each). The photometry was improved by stacking images from each night (from three to over thirty individual frames with seeing better than $1.5''$). We also attempted to assess the quiescent {\it I}-band brightness of the dwarf nova by stacking 122 frames with the best seeing ($<1''$). Although the depth of the resulting composite image reached $I\sim 23$~mag in empty sky regions, we were unable to resolve the nova owing to the proximity ($1.2''$) of a bright star ($I=16.6$~mag) and the glow of its image wings. Thus, we could only obtain an upper limit of $I>20.8$~mag on the nova brightness in quiescence.

Several images of Nova Cen 2009 were also collected immediately after the eruption, from 13 May to 26 May 2009. Because all these images were taken with the standard exposure time of 180~s, the nova is severely overexposed on each of them and thus its brightness could not be measured.

Post-eruption data were taken mainly as a part of the OGLE Galaxy Variability Survey\cite{udalski} with short exposure times from 25 to 30 s (123 images). They were supplemented with 119 deeper images (exposure time 180~s) obtained in 2010--2012 and March--May 2016. The latter were primarily used for the period analysis. 

\subsection{VLT data.}

Because V1213 Cen is located in the vicinity of transiting planet candidate OGLE-TR-167\cite{udalski2004}, it was serendipitously observed with the VIMOS instrument at the 8.2-m Very Large Telescope (VLT) on 11 April 2005 (under the program 075.C-0427(A), PI: D. Minniti). We stacked eleven images (with exposure time 15 s each) with the best seeing ($<0.63''$). The nova is clearly visible in that composite image. We used the DAOphot package\cite{stetson} to estimate its brightness at $V=24.4 \pm 0.4$ mag, which places an upper limit on the quiescent brightness of the dwarf nova (the star was not observed at that time by OGLE, thus we cannot assess whether it were fully quiescent).

\subsection{Orbital period.}

We used 107 deep $I$-band exposures acquired between 9 March and 3 May 2016 to measure the orbital period of the post-nova. There is a strong sinusoidal variability with a period of $5.0882 \pm 0.0017$ h, probably an orbital hump. In principle the orbital period might be twice as long. However, the absolute quiescent brightness of the nova ($M_V = +7.4 \pm 0.7$ mag) is inconsistent\cite{warner} with a $10.2$-h orbital period. (The secondary's mass would be larger than $1\ {\rm{ M_{\odot}}}$, which would imply $M_V < 5$ mag). Two additional periodicities at 4.1966 and 6.4605 h are day aliases of the orbital period. Assuming that the secondary fills its Roche lobe and follows the standard mass-radius relation for main-sequence stars\cite{warner}, its mass is $\approx 0.5\ {\rm M_{\odot}}$ and hence the $V$-band absolute magnitude $M_V \approx 8.9$ mag is in accordance with our observations. 

\subsection{Comparison with dwarf novae.}

There are some correlations between photometric features of dwarf nova outbursts\cite{warner}$^{\rm ,}$\cite{ot}. For example, the duration of outbursts is correlated with the orbital period\cite{ot}; the absolute magnitude during outbursts is correlated with the orbital period\cite{warner}$^{\rm ,}$\cite{pat11}; and the absolute magnitude in quiescence is correlated with the outburst frequency and orbital period\cite{warner}. Outbursts that have been observed before 2009 fullfil these correlations (see Extended Data Figs. 1c and 2). In the Extended Data Fig. 1, we compare photometric features (amplitude, inter-outburst time, outburst duration) of V1213 Cen with other dwarf novae\cite{ot} and see no differences between V1213 Cen and other dwarf novae. Moreover, the absolute brightness of V1213 Cen is also constistent with that of other dwarf novae of similar orbital period\cite{warner}$^{\rm ,}$\cite{pat11}$^{\rm ,}$\cite{rk}$^{\rm ,}$\cite{warner87}, see Extended Data Fig. 2.

\end{methods}

\newpage


\begin{figure}
\includegraphics[height=0.7\textheight]{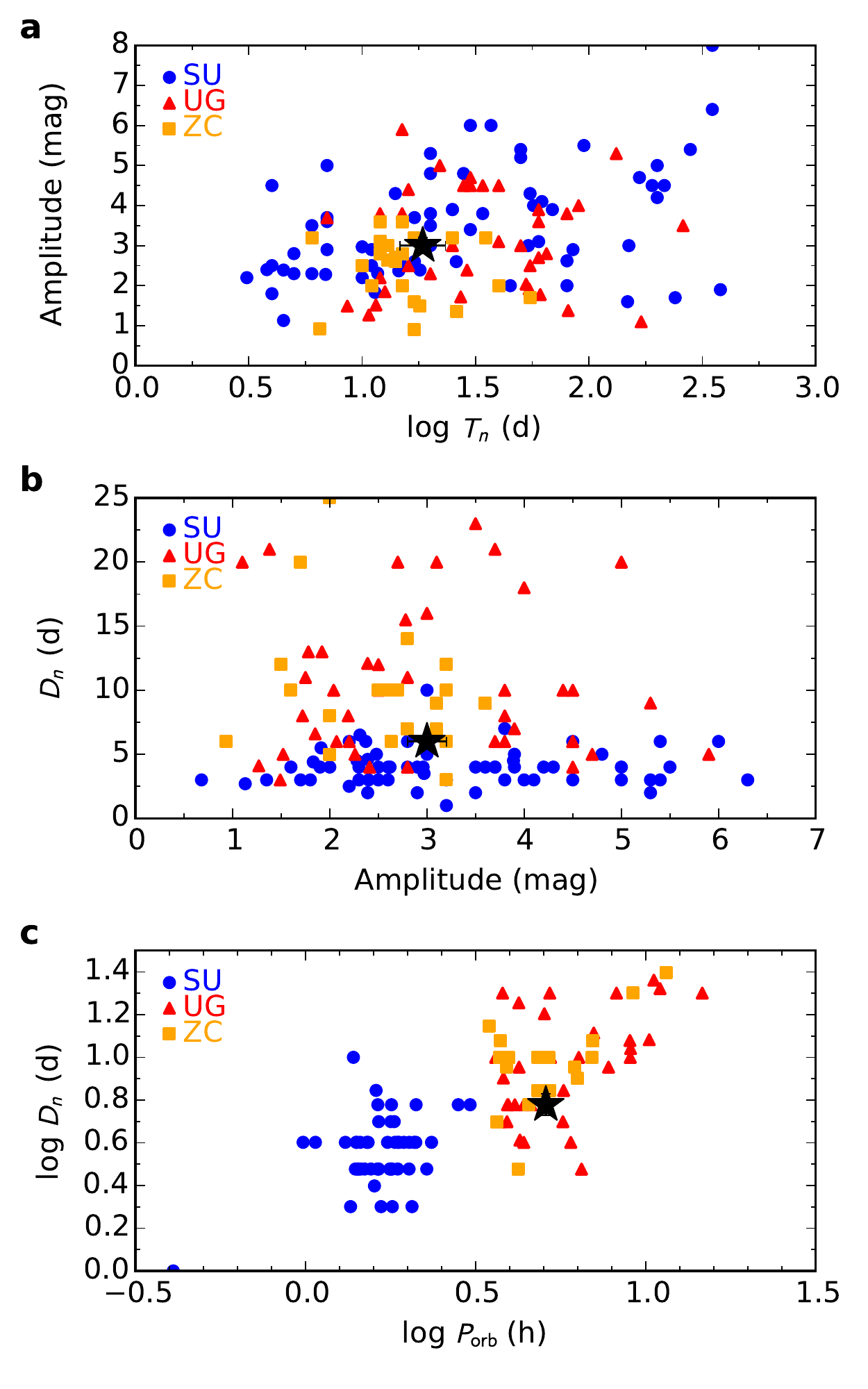} \\
\textbf{Extended Data Figure 1. Photometric features of dwarf novae.} a) $V$-band amplitude versus inter-outburst time $T_n$. b) Outburst duration $D_n$ versus $V$-band amplitude. c) Outburst duration $D_n$ versus orbital period $P_{\rm orb}$. The black asterisk shows V1213 Cen progenitor (with $1\sigma$ error bars). Blue dots, SU UMa-type dwarf novae; red triangles, U Gem-type dwarf novae; orange squares, Z Cam-type dwarf novae.
\end{figure}

\newpage

\begin{figure}
\includegraphics[width=\textwidth]{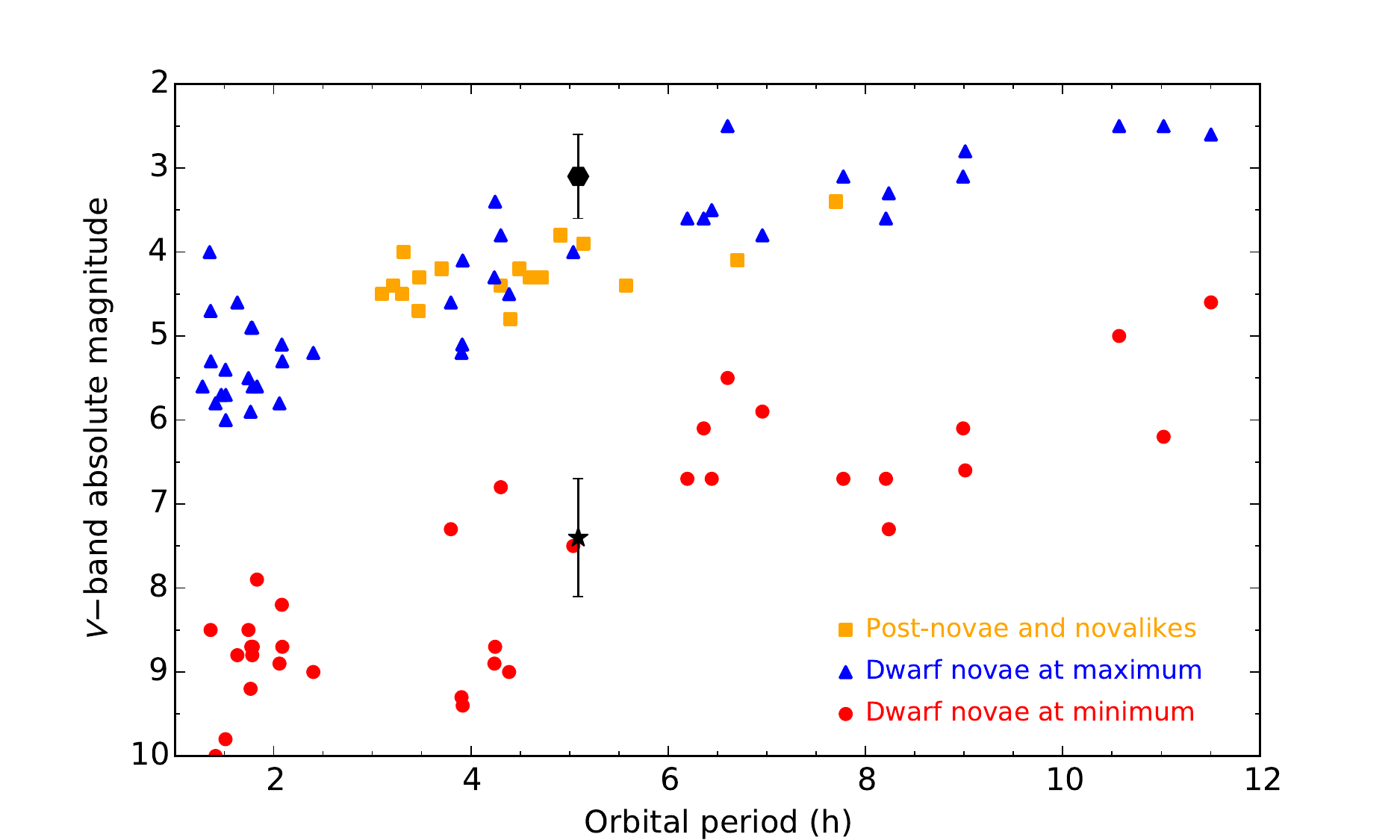}
\textbf{Extended Data Figure 2. Absolute magnitudes versus orbital periods for dwarf novae and post-novae.} The black asterisk shows the quiescent absolute magnitude of V1213 Cen progenitor (with $1\sigma$ error bars). The black hexagon indicates the current (2016) brightness of the post-nova.
\end{figure}

\end{document}